\documentclass[aps,pra,longbibliography,twocolumn]{revtex4-1}
\usepackage{amsmath,amssymb,amsthm,bm,bbm}
\usepackage{graphicx}
\usepackage{caption}
\usepackage{subcaption}
\usepackage{url}
\usepackage[usenames,dvipsnames]{color}
\usepackage{hyperref}
\usepackage{epstopdf}

\begin{document}
\newcommand{\cs}[1]{}

\title{Correlation Between Student Collaboration Network Centrality and Academic Performance}
\author{David L. Vargas}
\author{Ariel M. Bridgeman}
\author{David R. Schmidt}
\author{Patrick B. Kohl}
\author{Bethany R. Wilcox}
\author{Lincoln D. Carr}
\affiliation{Department of Physics, Colorado School of Mines, Golden, CO 80401, 
USA}
\date{\today}

\begin{abstract}
We compute nodal centrality measures on the collaboration networks of students enrolled in three upper-division physics courses, usually taken sequentially, at the Colorado School of Mines.  These are complex networks in which links between students indicate assistance with homework.  The courses included in the study are intermediate Classical Mechanics, introductory Quantum Mechanics, and intermediate Electromagnetism.  By correlating these nodal centrality measures with students' scores on homework and exams, we find four centrality measures that correlate significantly with students' homework scores in all three courses: in-strength, out-strength, closeness centrality, and harmonic centrality.  These correlations suggest that students who not only collaborate often, but also collaborate significantly with many different people tend to achieve higher grades.  Centrality measures between simultaneous collaboration networks (analytical vs. numerical homework collaboration) composed of the same students also correlate with each other, suggesting that students' collaboration strategies remain relatively stable when presented with homework assignments targeting different skills.  Additionally, we correlate centrality measures between collaboration networks from different courses and find that the four centrality measures with the strongest relationship to students' homework scores are also the most stable measures across networks involving different courses.  Correlations of centrality measures with exam scores were generally smaller than the correlations with homework scores, though this finding varied across courses.   
\end{abstract}

\maketitle

\section{Introduction}
Physics education research has enjoyed a great deal of success in identifying and clarifying students' difficulties with physics concepts,  developing problem solving methods, and structuring the  knowledge that is taught to students \cite{McDermott1999,McDermott1984,Heuvelen1991,Reif1986, Redish1999}.  Such studies have allowed researchers to make quantitative statements about the presence and persistence of students' difficulties, in contrast to historical physics education that relied on anecdotal information \cite{McDermott1984}.  In recent years, physics education researchers have begun taking advantage of another powerful tool for quantitative analysis in the social sciences known as complex network theory \cite{Borgatti2009, Newman2003}, the use of which is often referred to as social network analysis (SNA).  SNA has a significant history of use in educational research generally \cite{wasserman1994,Newman2003,Grunspan2014}; however, the use of SNA has only recently begun to gain traction within the physics education research community (e.g., Refs.\ \cite{Bodin2012,Bruun2013}).  

In this paper, we use SNA to study self-reported student collaboration and its potential relation to student performance.  We examine how collaboration between students evolves between semesters and how nodal centrality measures correlate with homework vs. exam grades.  Furthermore, we compute the differences in correlation strengths between our measures, allowing us to quantify which measures are most strongly related to student grades.  Finally, we also compare the network centrality of students between the collaboration networks of different types of homework assignments within a single course, allowing us to assess the similarity of roles adopted by students in response to assignments of differing nature.  

Complex network measures provide succinct summaries of the order present in complex networks.  Often such measures are aggregate summaries of the entire structure of a network and are useful because the structure of connectivity can determine the efficiency of processes taking place in the network as observed in social, neural, communication, and transportation networks \cite{Newman2003,Latora2001}.  Furthermore, the nodal centrality measures we review in Sec.\ \ref{sec:cna} tell us how well connected students are in the context of their homework collaboration networks \cite{Freeman1977,Freeman1979}.  For example, a simple measure of how well a student is connected to other students in the network is encapsulated by their \textit{out-strength}.  Out-strength is simply the number of peers a student helps with homework and, thus, is one measure of the influence of student in a collaboration network.  Parallel to out-strength is \textit{in-strength}, the number of students that help a particular student with homework.  While out-strength can be thought of as a coarse measure of the influence of a student on the collaboration network, in-strength is a measure of how a student gathers information from different parts of the network.  Other more subtle measures of a student's connections within the network include \textit{closeness centrality}, which looks at the  ``distance'' between students (i.e., if information is to pass from student $i$ to student $j$, how many other students must it go through first), as well as \textit{betweenness centrality}, which looks at the degree to which a particular student can control the flow of information between other students.  These centrality measures are discussed in detail in Sec.\ \ref{sec:cna}.  The variety of complex network measures we consider provide us with different perspectives of the students forming our collaboration networks.  The aggregate correlation of the centrality measures with student grades provide insight into how strongly and in what manner different collaboration patterns are central to the educational process.

While the application of complex network-based methods to student networks formed within the physics classroom are rare, there is a growing body of work in this area.  In particular, network analysis has been consistently applied to investigations of student persistence \cite{Brewe2012,Zwolak2017,Zwolak2018,Forsman2014}.  The use of network analysis in the context of understanding student persistence is motivated by the idea that a student's decision to persist (or not) within a particular major is impacted by their integration within, and interaction with, their academic and social communities \cite{Zwolak2017}.   Forsman \emph{et al.} reported on students' in-class social and academic interactions.  They showed that these social and academic networks had distinct connection patterns indicating that the underlying processes governing the formation of these networks is different.  They then argued that both networks need to be considered when addressing student persistence.  Zwolak \emph{et al.} \cite{Zwolak2017} created networks based on students' in-class interactions within a highly interactive introductory course and found that certain network centraility measures correlated significantly with students' persistence into the next course in the sequence.  This correlation held even when controlling for the impact of a student's grade.  Zwolak \emph{et al.} \cite{Zwolak2018} later built on this work by incorporating an additional network based on students' interactions outside of class.  They found that for middle-performing students, out-of-class centrality measures dominated in terms of predicting students' persistence to the next course in the sequence.  

Network analysis has also been utilized in investigations not directly tied to student persistence.  Dou \emph{et al.} \cite{Dou2016} reported correlations between network centrality measures and changes in students' self-efficacy over the course of an introductory physics course taught using modeling instruction.  Brunn and Brewe \cite{Bruun2013} used network centrality measures calculated based on students' interactions in an introductory physics course to predict their grades in a future course.  The use of network analysis in physics education research has also been extended beyond social networks.  For example, Bodin \cite{Bodin2012} applied network analysis to visualize connections between students' epistemic ideas when solving physics problems involving simulations and modeling tasks, and Brewe \emph{et al.} applied network analysis to characterize students' responses to an introductory conceptual assessment.  


The current study is distinct from the work described above in several ways.  The majority of the prior work focuses on students' in-class interactions in the context of an introductory physics course utilizing a highly interactive curriculum.  The current work focuses on three upper-division courses all taught with a mixture of traditional lecture punctuated by the use of interactive techniques.  Additionally, the current work deals with students' out-of-class collaborations on homework assignments; thus, these interactions are driven almost entirely by the students without the significant pressure to collaborate usually associated with an interactive classroom environment.  Using a wide variety of complex network measures, we obtain detailed information about the role of different collaboration strategies in different types of problem sets and on exams.  This study is a step towards responding to multiple calls to take advantage of the analysis power of complexity science within physics education research \cite{Forsman2014,Zwolak2018}.  

Here, we build off the work described above by addressing the following set of questions.  Do well-connected students have good grades?  Does access to the reasoning of many of their peers better equip students to complete homework assignments, or does excessive participation in a collaboration network stifle the ability of a student to perform well on their own work?  Do the benefits of collaboration extend to exams, where a student does not have access to their collaborators?  How stable are these measures of collaboration in different contexts?  That is, do students tend to take on different roles in response to different types of assignments or different subject-matters, or are students' collaboration strategies static?  

\section{Data Collection}
The Colorado School of Mines (Mines) is a public research university in Golden, Colorado.  The university, which has close to six thousand undergraduate and graduate students, focuses on engineering and the applied sciences.  Additionally, it is one of very few institutions that awards more than 50 Physics Bachelors per year, placing it in the top ten of all Ph.D. granting departments in the U.S. \cite{AIP}.  The physics department has research focus areas in condensed matter, subatomic, optical, renewable energy, theoretical, and computational physics.

\begin{figure}
\includegraphics[page=1, width=.85\linewidth]{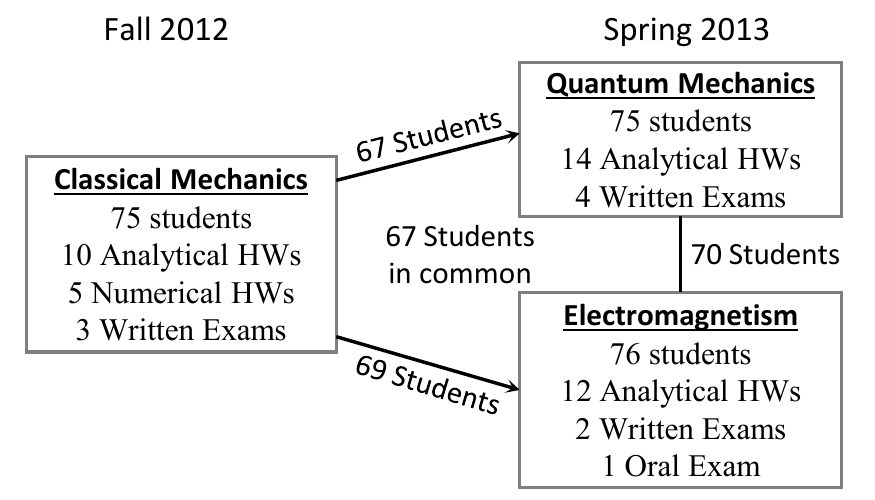}
\caption[Course information for Classical Mechanics, Quantum Mechanics, and Electromagnetism.]{\textit{Course information for Classical Mechanics, Quantum Mechanics, and Electromagnetism.} Lines connecting two courses indicate the number of students common to both courses.  There were 67 students enrolled in all three courses.}
\label{fig:course_info}
\end{figure}

The data for our networks was collected over two semesters: Fall 2012 and Spring 2013.  Students in their junior year in engineering physics at Mines take Classical Mechanics during the Fall semester and both Quantum Mechanics and Electromagnetism during the Spring semester. The course in Classical Mechanics covers Lagrangian and Hamiltonian mechanics.  The Electromagnetism course is the first course in a two-course sequence, and covers electrostatics and magnetostatics, including the appropriate Maxwell equations, boundary conditions, and treatments of free and bound sources.  The course in Quantum Mechanics introduces the formalism of Quantum Mechanics (e.g., the solutions of a particle in a box, scattering from a potential well, etc.).  We summarize the homework, exam, and enrollment information for the three courses in Fig.~\ref{fig:course_info}.  Note that the longitudinal nature of our data means that many students in our data set appear in all three courses.  Prior to their junior year at Mines, physics majors are encouraged to collaborate in a physics studio setting, a setting in which students work in groups of three to complete homework-like assignments and labs.  Additionally, the summer before their junior year, Mines physics majors participate in a physics field session in which groups of ten students move between sections on computing, vacuum systems, machining, and lasers.  In all of these sections students are encouraged to collaborate, and in some of them students are split into groups of three to complete assignments.  Thus, collaboration is an explicit part of Mines' lower-division program already, and is strongly encouraged.

There were two forms of data collection: paper forms during Fall 2012 and electronic spreadsheets during Spring 2013.  In the paper form of data collection, students were provided with a form for each assignment in which they were to list any students they helped or received help from for each assigned problem.  In the electronic form of data collection, students provided the same data by entering the names of their collaborators into question/answer boxes on the learning management system Blackboard \cite{Blackboard}.  In both cases, students were required to complete the surveys described above in order to receive credit for their homework assignments.  This policy incentivized survey completion and ensured a nearly complete set of data.  Student names were then replaced with a set of randomly generated three letter codes to anonymize the data prior to analysis.

The data from the surveys above was compared with student grades in the three courses (Fig.\ \ref{fig:course_info}).  The course in Classical Mechanics had ten analytical homework assignments, five numerical homework assignments, and three written exams.  For numerical assignments, students were asked to simulate various physical scenarios using Mathematica.  For the course in Classical Mechanics, we computed three measures of a students' performance: the sum of their analytical homework grades, the sum of their numerical homework grades, and the sum of their exam grades.  For the courses in Quantum Mechanics and Electromagnetism, all homework assignments were analytical (see Fig.\ \ref{fig:course_info}), and we measured a students' performance by the sum of their homework grades and the sum of their exam grades.  It is important to note that in Classical Mechanics, the teaching assistants graded exams with subsequent review by the instructor.  In Quantum Mechanics and Electromagnetism, the instructors graded all exams themselves.  Finally, the instructors of Quantum Mechanics and Classical Mechanics both applied curves to exam grades, whereas the instructor of Electromagnetism did not. 

\section{Methodology for Converting Data into Networks}
From the data collected in the surveys above, we constructed directed and weighted networks for each course and type of homework assignment using the following procedure. A network is a collection of nodes and links.  Nodes are any object that can be connected to any other object by some relation and links are the connections between nodes.  In our networks, nodes correspond to students, and a link corresponds to an interaction between the students consisting of providing or receiving assistance.  Note that this is not a symmetric relation; if $i$ helped $j$, it does not imply that $j$ helped $i$.  This is the defining feature of a directed network; its connections are asymmetrical.  For directed networks, one says that a link goes from node $i$ to node $j$ to indicate the direction of the link.  For the $k^{\textrm{th}}$ network a link is placed from node $i$ to node $j$ if and only if (iff) student $i$ helped student $j$ with homework assignment $k$.  Summarizing our network in terms of the entries of an adjacency matrix,
\begin{equation}
  \label{eq:directedadjmatrix}
  A_{ij}^k = 
\begin{cases}
1 \textrm{ iff student $i$ helped student $j$}\\
\textrm{ \hspace{2mm}with homework assignment $k$} \\
0  \textrm{ otherwise.}
\end{cases}  
\end{equation}

\begin{figure*}
\begin{subfigure}[t]{0.5\textwidth}
\includegraphics[page=1, width=\linewidth]{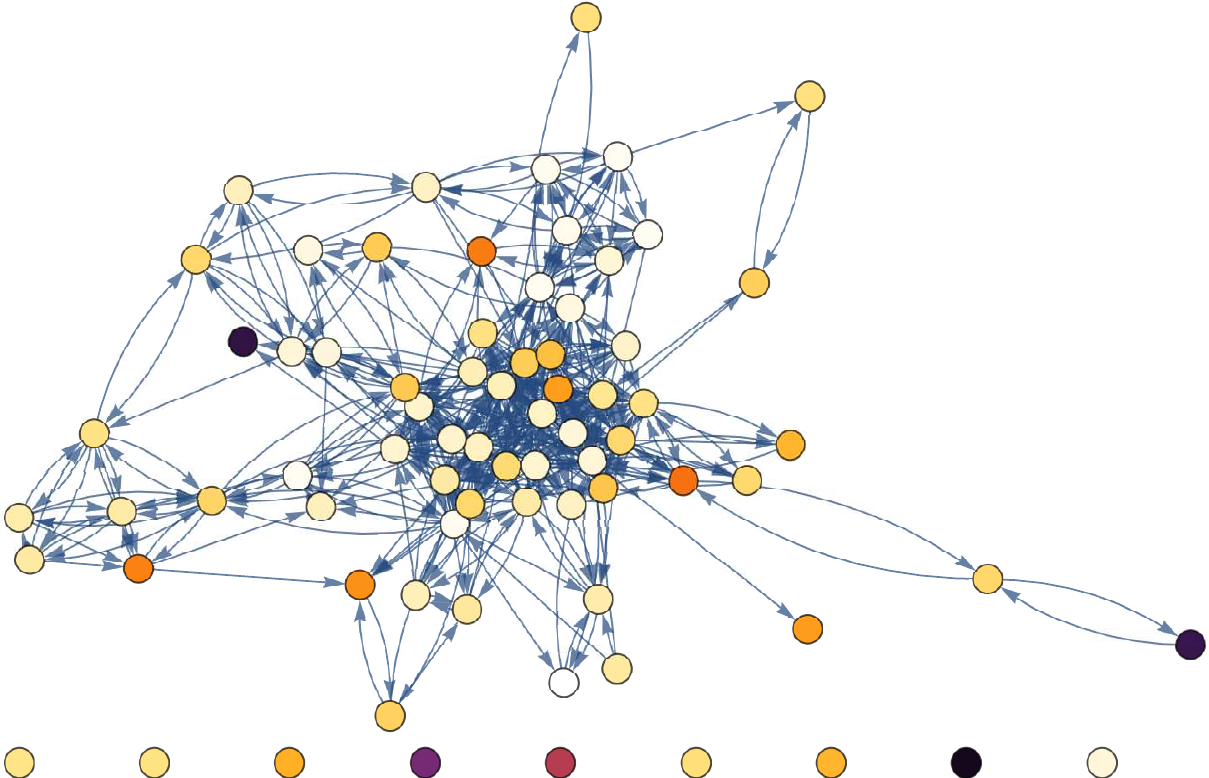}
\caption{Analytical Assignments Network for Classical Mechanics }
\end{subfigure}%
\begin{subfigure}[t]{0.5\textwidth}
\includegraphics[page=1, width=\linewidth]{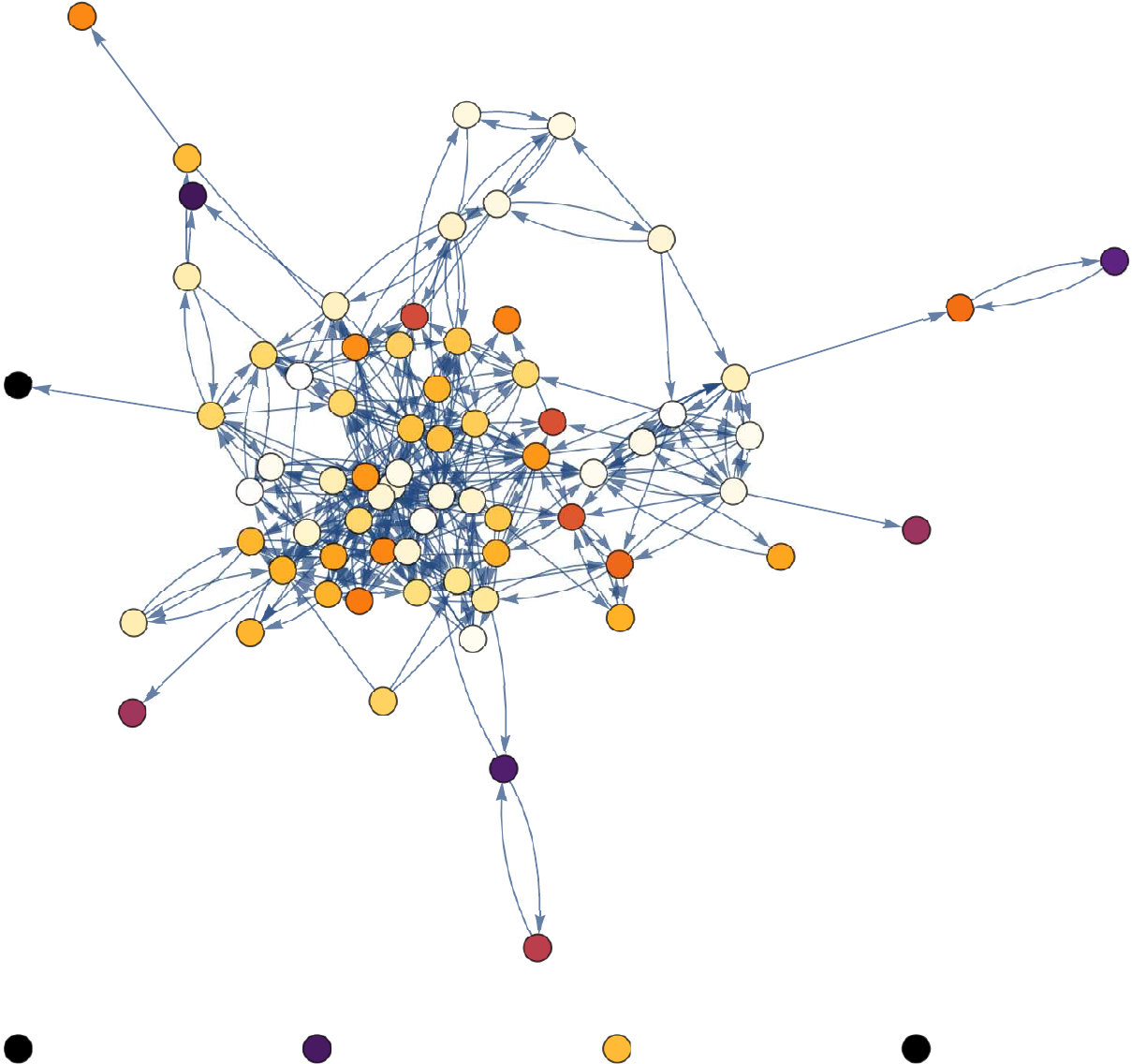}
\caption{Numerical Assignments Network for Classical Mechanics}
\end{subfigure}
\begin{subfigure}[t]{0.4\textwidth}
\includegraphics[page=1, width=\linewidth]{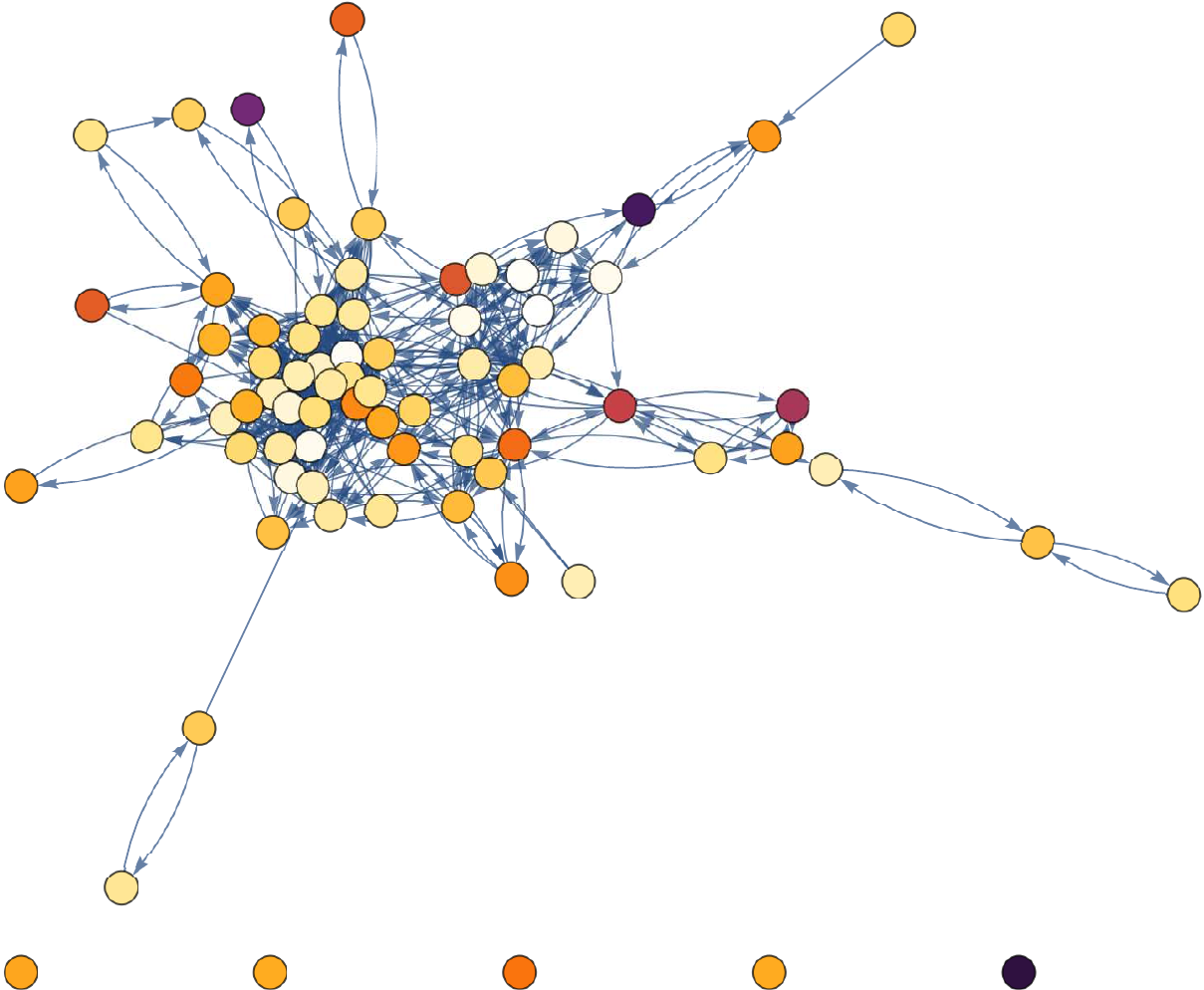}
\caption{Assignments Network for Quantum Mechanics}
\end{subfigure}%
\begin{subfigure}[t]{0.4\textwidth}
\includegraphics[page=1, width=\linewidth]{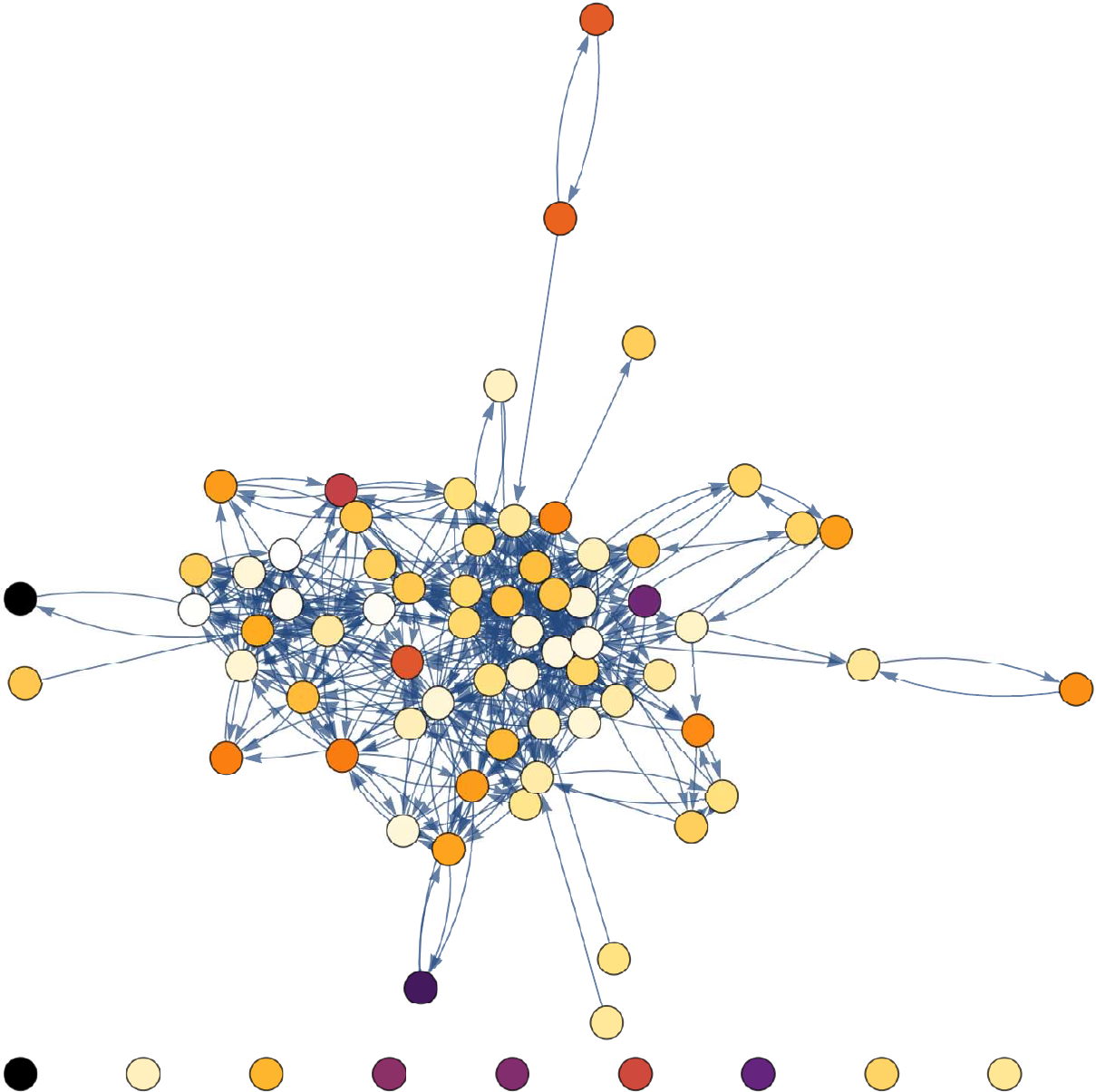}
\caption{Assignments Network for Electromagnetism}
\end{subfigure}%
\begin{subfigure}[t]{0.1\textwidth}
\includegraphics[page=1, width=\linewidth]{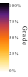}
\end{subfigure}
\caption{\textit{Student collaboration networks for three upper level physics courses.}  Weighted student collaboration networks constructed from surveys given to students in three upper-division courses: Classical Mechanics, Quantum Mechanics, and Electromagnetism.  Nodes correspond to students and the direction of each arrow indicates the direction of assistance on homework assignments.  The color of a node indicates the grade of a student on homework assignments.  Although we do not normalize grades in our analysis we present grades as a percentage here to illustrate multiple courses simultaneously.}\label{fig:collaboration-networks}
\end{figure*}

However, we found that it was necessary to resolve discrepancies in the reports provided by students.  For example, student $i$ may claim that they helped student $j$, but student $j$'s survey indicates that they did not receive help from student $i$.  These discrepancies may be due to forgetfulness or conflicting perceptions of interactions \cite{Bruun2013}.  To resolve the discrepancies in student reports, we employed a \textit{Maximal} discrepancy resolving technique \cite{AAPT,PERC}.  Applying an element-wise logical OR to the adjacency matrices created from each student's survey we made a final network.  That is, every reported interaction is considered to have happened even if one student does not report it.  Discrepancies can occur in either direction of an interaction, and $A_{ij}$ is resolved separately from $A_{ji}$.  Other discrepancy sorting cases were investigated but yielded quite sparse adjacency matrices \cite{AAPT,PERC}.  For each course, we then compute a weighted adjacency matrix by summing the adjacency matrices corresponding to the homework assignments in that course,
\begin{equation}
  \label{eq:weighted-network}
  A_{ij} = \sum_{k=1}^{N_{\textrm{HW}}} A_{ij}^k\,.
\end{equation}
Where $N_{\textrm{HW}}$ is the number of homework assignments in the relevant course.  Thus if two students $i$ and $j$ collaborated frequently on homework assignments they will have a heavily weighted connection in one of the weighted networks depicted in Fig.~\ref{fig:collaboration-networks}.  In Fig.~\ref{fig:collaboration-networks}, nodes are indicated by circles and the links connecting nodes are indicated by the arrows between nodes.  The direction of the arrow indicates the direction of assistance.  For the course in Classical Mechanics, we construct networks for the numerical homework assignments and the analytical homework assignments separately.  This procedure results in two networks for the course in Classical Mechanics: a network constructed from the collaboration networks on analytical assignments $A^{\textrm{CM}}_A$, and a network constructed from the collaboration networks on numerical assignments $A^{\textrm{CM}}_N$.   We denote the  network for the course in Quantum Mechanics by $A^{\textrm{QM}}$, and the network for the course in Electromagnetism as $A^{\textrm{EM}}$.


\section{Complex Network Analysis}
\label{sec:cna}
Using the NetworkX network analysis software \cite{hagberg-2008}, as well as some of our own independently developed network analysis code, we study the networks described in the previous section by computing various nodal centrality measures and other measures of the structure of a node's connections.  We then compute the correlation between these measures and different estimators of student performance.  Nodal centrality measures are measures of a node's importance to the structure of the network.  As they are quantitative measures of each student's role in the collaboration network, nodal centrality measures are ideal for our study.  Our selection of standard complex network measures includes out-strength, in-strength, out-disparity, in-disparity,  local clustering, closeness centrality, harmonic centrality, and betweenness centrality.  The local clustering coefficient is only defined on undirected networks.  Before computing the local clustering coefficient we first convert our directed networks into undirected networks as such that, $A^{\textrm{undirected}}_{ij} = \max\left(A_{ij},A_{ji}\right)$.

The \textit{out-strength} of a node is the sum of its outgoing connections to other nodes and is defined as
\begin{equation}
  \label{eq:sout}
  s^{\textrm{out}}_i  = \sum_{j=1}^L A_{ij}\,,
\end{equation}
where $L$ represents the number of nodes (i.e., students) in the network.  A node can have high out-strength if it has outgoing connections to many other nodes, or if it has strong connections to only a few other nodes.  Stated simply, students who help many of their peers and students who frequently help a smaller set of peers both can have a high out-strength.  The \textit{in-strength} is similarly defined and distinguishes a node by the number of incoming connections
\begin{equation}
  \label{eq:sin}
s^{\textrm{in}}_i  = \sum_{j=1}^L A^T_{ij}\,,
\end{equation}
or the number of instances in which a student received help.  We also study the \textit{net out-strength} 
\begin{equation}
  \label{eq:snet}
s^{\textrm{net}}_i = s^{\textrm{out}}_i - s^{\textrm{in}}_i\,.
\end{equation}  
Students with high net out-strength correspond to students that help many other students but are not helped by many students.

The \textit{out-disparity} of a node's connections is a measure of the non-uniformity of the outgoing connection strengths.  If a node has a single strong connection in addition to other, much weaker connections, the node has high out-disparity.  If the connection strengths of a node are all approximately equal strength, then it has a low out-disparity. Out-disparity is defined as \cite{Boccaletti2006, barabasi2004}
\begin{equation}\label{eq:disparity}
Y^{\textrm{out}}_i \equiv  \frac{1}{\left(s^{\textrm{out}}_i\right)^2}\sum_{j=1}^L
\left(A_{ij}\right)^2 = \frac{\sum_{j=1}^L
\left(A_{ij}\right)^2}{\left(\sum_{j=1}
^L A_{ij}\right)^2}\, .
\end{equation}
Nodes with high disparity correspond to students that collaborate with certain nearest neighbors much more often than they collaborate with other nearest neighbors. Nodes with low disparity correspond to students that collaborate equally with all students that they collaborate with.  Analogously, \textit{in-disparity} measures the non-uniformity of the incoming connection strengths.  To compute $Y^{\textrm{in}}_i$ one makes the substitution $A\rightarrow A^{T}$ in Eq.~\eqref{eq:disparity}, resulting in
\begin{equation}
\label{eq:indisparity}
Y^{\textrm{in}}_i \equiv  \frac{1}{\left(s^{\textrm{in}}_i\right)^2}\sum_{j=1}^L
\left(A^{T}_{ij}\right)^2 = \frac{\sum_{j=1}^L
\left(A^{T}_{ij}\right)^2}{\left(\sum_{j=1}
^L A^{T}_{ij}\right)^2}\,.
\end{equation}
Note that out-disparity can only be defined for student's with $s^{\textrm{out}}_i>0$; thus students with $s^{\textrm{out}}_i=0$ are not included in correlations involving out-disparity.  The same holds for in-disparity for student with $s^{\textrm{in}}_i=0$. 

The \textit{local clustering coefficient} is a measure of the transitivity of connections of individual nodes, that is, the likelihood that $a$ is connected to $c$, given that $a$ is connected to $b$ and $b$ is connected to $c$.  The local clustering coefficient is defined as 
\begin{equation}
  \label{eq:localclustering}
  c^L_i \equiv \frac{T(i)}{k_i(k_i-1)},
\end{equation}
where $T(i)$ is the number of existing triangles in which node $i$ is a vertex, and $k_i$ is the degree of node $i$.   Effectively, this provides the fraction of all possible triangles through node $i$ that actually exist.  Nodes with low local clustering correspond to students whose collaborators do not tend to collaborate with each other.  Nodes with high local clustering correspond to students whose collaborators frequently collaborate with each other, such as in tight-knit study groups.

In a weighted network one can define a distance between any pair of nearest-neighbor nodes.  For our analysis, we define the distance between nearest neighbors $i$ and $j$ to be the inverse of the weight connecting them 
\begin{equation}
D_{ij} = \frac{1}{A_{ij}}\,.
\end{equation}
If nodes $i$ and $j$ are not directly connected by a link then $D_{ij} = \infty$.  This definition of the distance between nearest-neighbor nodes is then used to define the shortest-path distance between any two nodes $d_{ij}$.  A path connecting node $i$ to node $j$ is a sequence of links along which one may walk to traverse the network from node $i$ to node $j$ when one walks along links in the direction of the link.  The shortest-path distance between two nodes is the sum of the nearest-neighbor weights $D_{ij}$ along the shortest path connecting two nodes, that is,
\begin{equation}
\label{eq:shortest-path}
d_{ij} = \underset{P}{\textrm{min}} \sum_{(l,k) \in P} D_{lk}\,,
\end{equation}
where $P$ is a path connecting node $i$ to node $j$.

\textit{Closeness centrality} is a measure of how close a node is on average to other nodes when one must travel along directed links in the direction of the link. Closeness centrality is defined as
\begin{equation}
  \label{eq:closeness_centrality}
  c^C_i = \frac{n-1}{|A| - 1}\frac{1}{\sum_{j \neq i} d_{ij}},
\end{equation}
where $n$ is the number of nodes reachable from node $i$, and $|A|$ is the number of nodes in the network defined by the adjacency matrix $A$ \cite{hagberg-2008, Freeman1979}.  Reachable means that one can travel from node $i$ to node $j$ by walking along links in the direction of the link.  Any nodes that are not reachable from node $i$ are neglected in the sum of Eq. \eqref{eq:closeness_centrality}.    In the context of social networks, closeness centrality can be thought of as a measure of independence as described in \cite{Freeman1979}.  This is because a node with a large closeness centrality does not have to rely on any one or two other nodes to transmit messages across the network \cite{Freeman1979}.  In the context of weighted student collaboration networks, closeness centrality is a measure of both the frequency with which a student assists others and how widely a student collaborates.

\textit{Harmonic centrality} is also a measure of how close a node is to other nodes in the network when one must travel along directed links in the direction of the link.  Harmonic centrality is defined as
\begin{equation}
  \label{eq:harmonic_centrality}
  c^{H}_i = \sum_{j \neq i} \frac{1}{d_{ij}},
\end{equation}
where $d_{ij}$ is the shortest path distance from node $i$ to $j$ \cite{hagberg-2008, Boldi2014}.  Harmonic centrality has a similar definition to closeness centrality, both being defined in terms of the inverse distances between nodes.  The intuition for the two measures is the same.  Nodes that are close to other nodes are more central as measured by closeness centrality and harmonic centrality.  However, when computing harmonic centrality, if node $j$ is not reachable from node $i$, then the distance between the two nodes is set to $d_{ij} = \infty$.  The corresponding term in the sum is then set to zero, $1/d_{ij} = 1/\infty \equiv 0$.  This may be preferable to the procedure used to calculate closeness centrality as this procedure has been shown to introduce a bias towards nodes in small components because it does not take into account nodes that are not reachable by the node of interest \cite{Boldi2014}.


\textit{Betweenness centrality} is measure of how important a node is as a go-between for message transmission between nodes in a network, assuming that information travels along paths of shortest distance \cite{Newmanbook}.  Betweenness centrality is defined as
\begin{equation}
  \label{eq:betweenness_centrality}
  c^{B}_i = \sum_{j,k \in V} \frac{\sigma(j, k | i)}{\sigma(j,k)},
\end{equation}
where $\sigma(j,k|i)$ is the number of shortest paths from node $j$ to node $k$ that pass through node $i$ and where $\sigma(j,k)$ is the number of shortest paths from node $j$ to node $k$ \cite{hagberg-2008, Freeman1977}.  Nodes with high betweenness centrality correspond to students with the most control over information transfer throughout the network.  Therefore, the weight of the links in our networks do not modify betweenness centrality directly, but do indirectly contribute through the path lengths.



\section{Results}
\begin{figure*}
\begin{subfigure}[t]{0.5\textwidth}
\includegraphics[page=1, width=\linewidth]{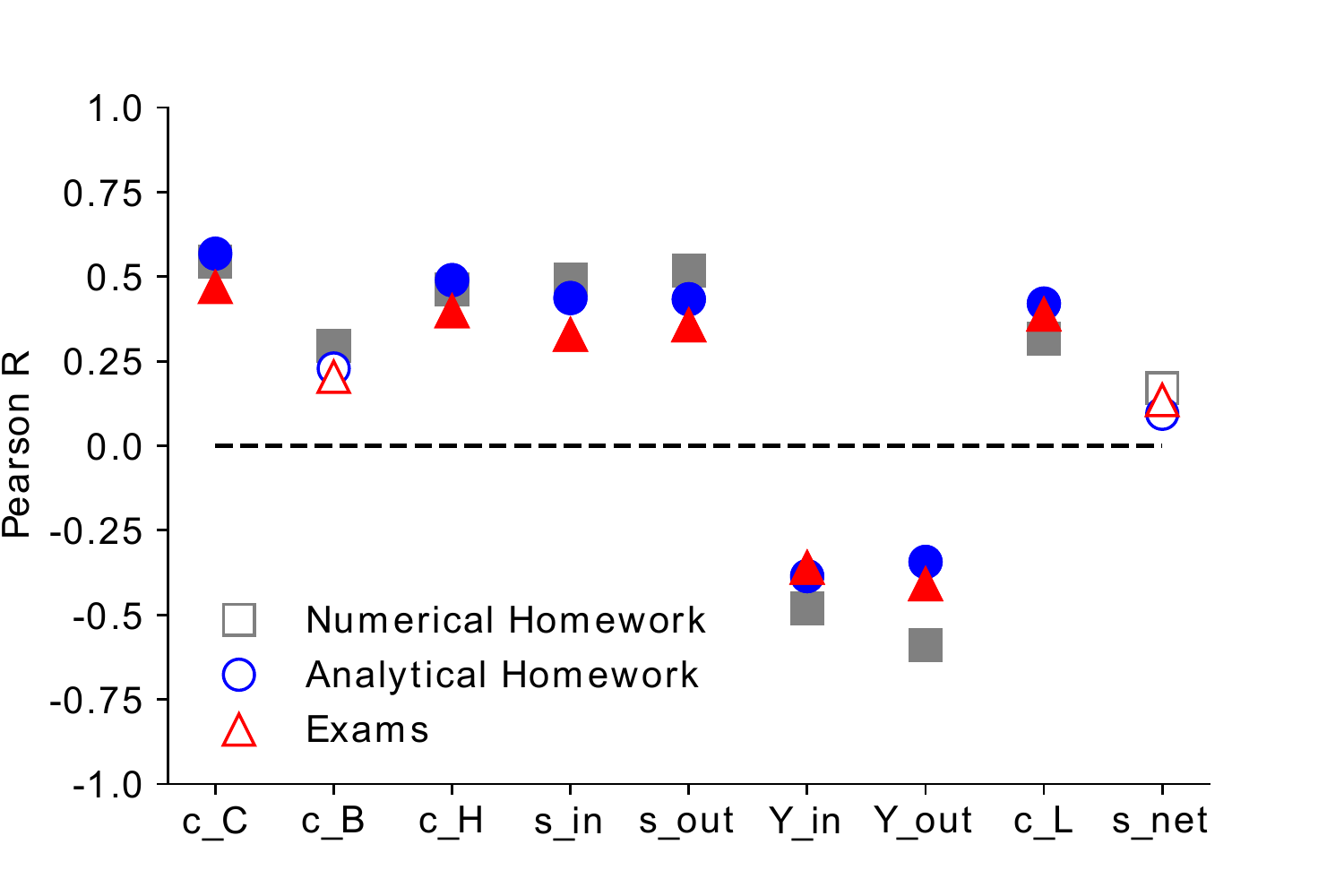}
\caption{Classical Mechanics Analytical Network}
\end{subfigure}%
\begin{subfigure}[t]{0.5\textwidth}
\includegraphics[page=1, width=\linewidth]{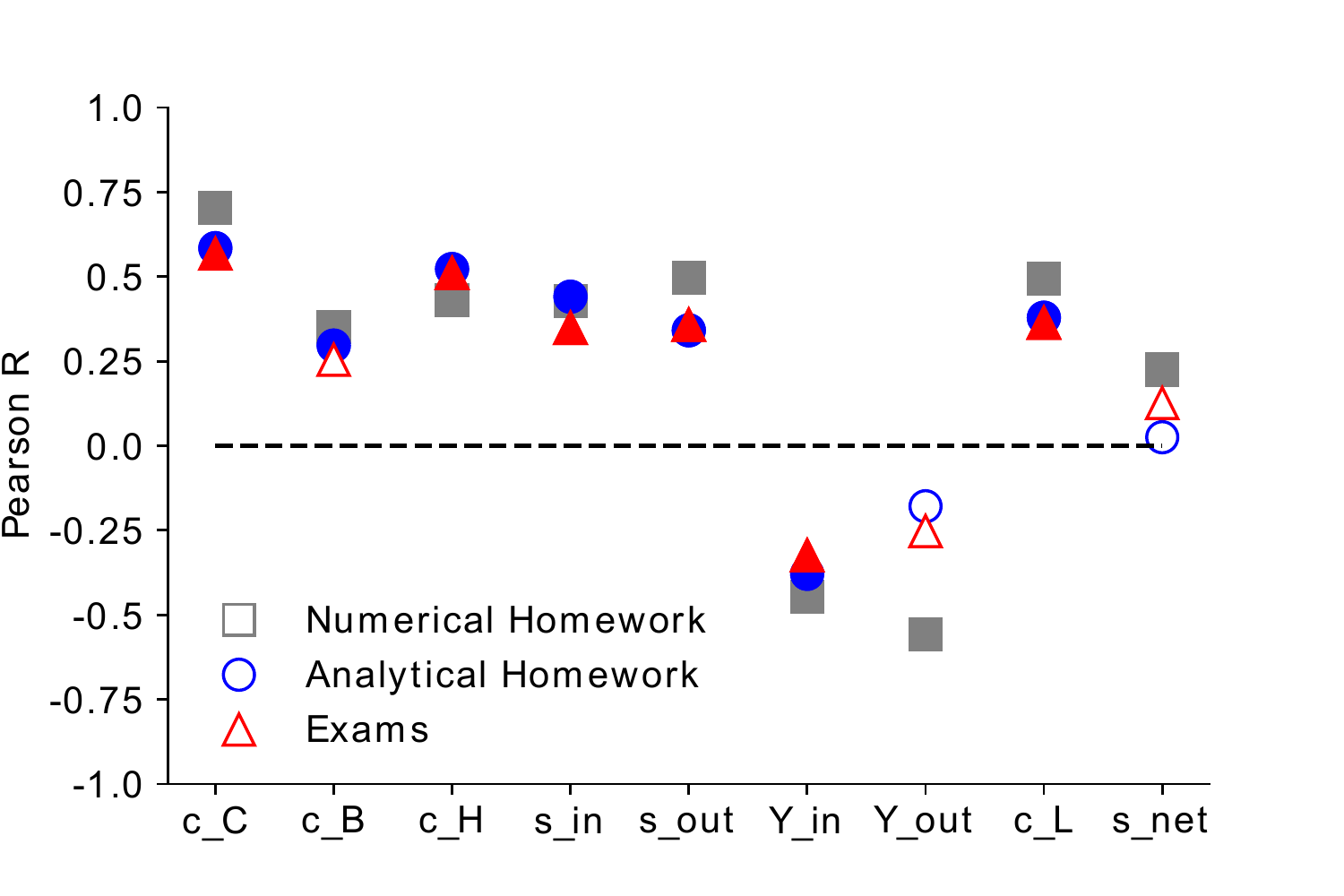}
\caption{Classical Mechanics Numerical Network}
\end{subfigure}
\begin{subfigure}[t]{0.5\textwidth}
\includegraphics[page=1, width=\linewidth]{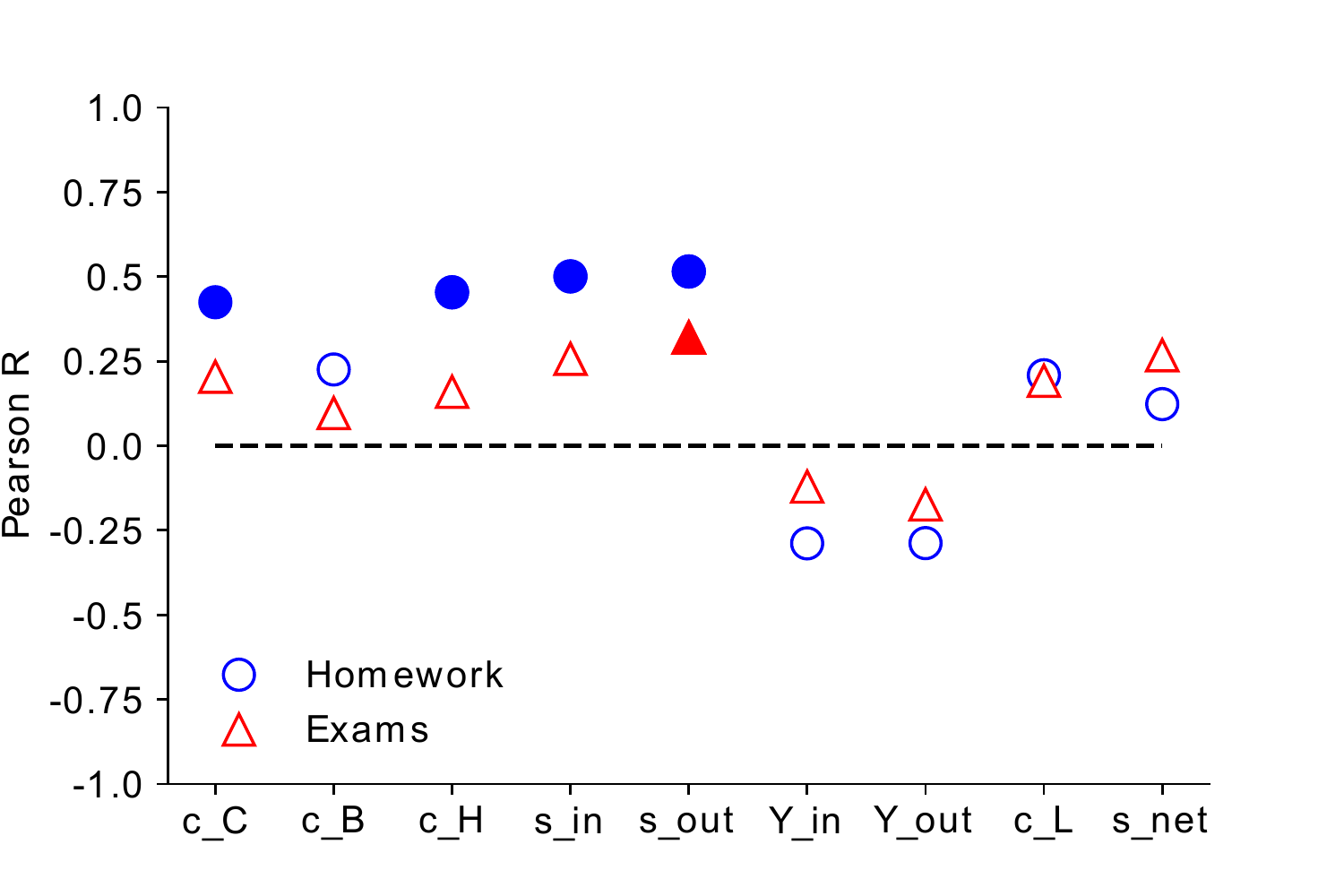}
\caption{Quantum Mechanics Network}
\end{subfigure}%
\begin{subfigure}[t]{0.5\textwidth}
\includegraphics[page=1, width=\linewidth]{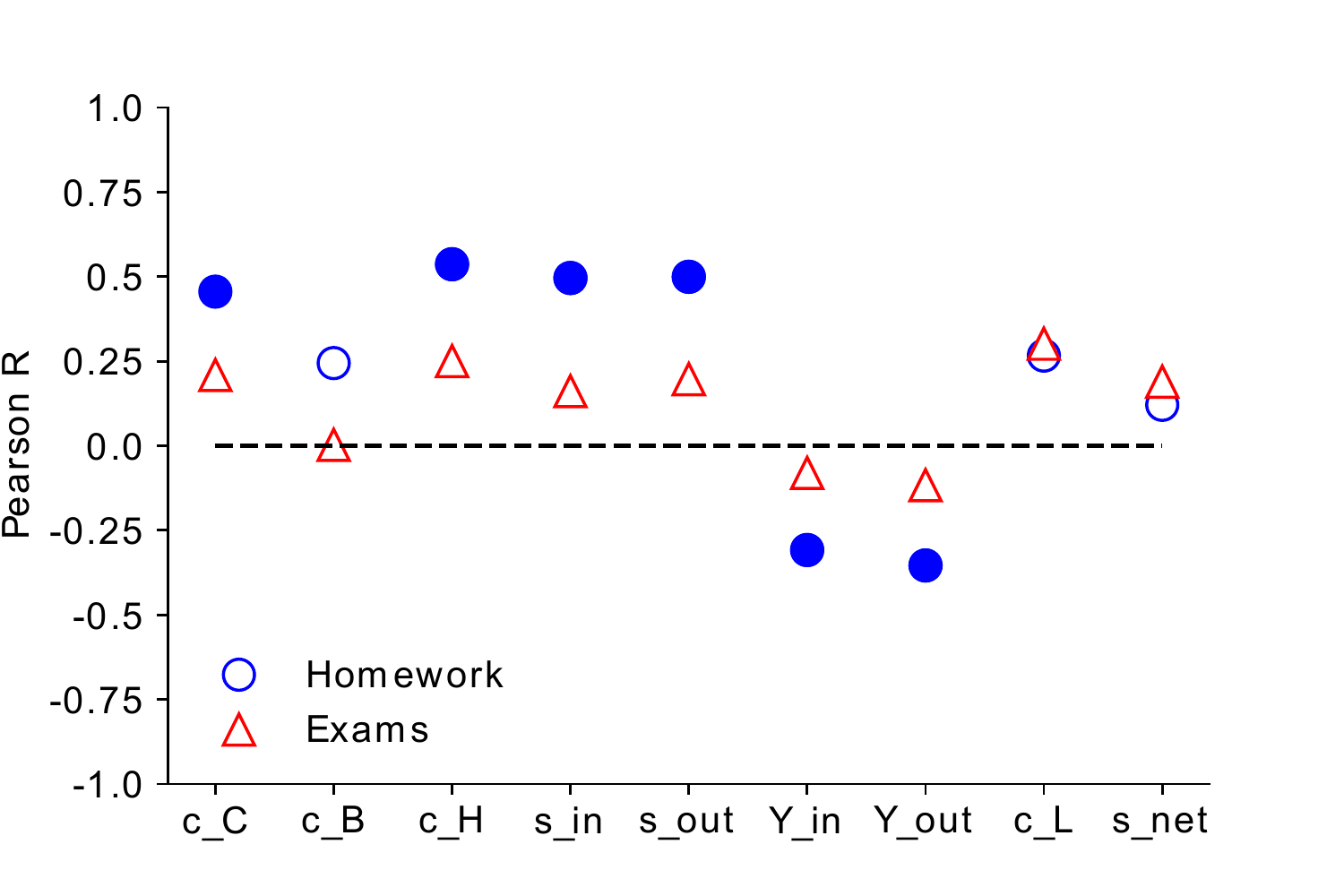}
\caption{Electromagnetism Network}
\end{subfigure}
\caption[Correlation of nodal centrality measures with student grades in three upper-division physics courses.]{\textit{Correlation of nodal centrality measures with student grades in three upper level physics courses.} Correlation of complex network measures with student grades for three courses: Classical Mechanics, Quantum Mechanics, and Electromagnetism. Filled markers indicate correlation coefficients that are statistically significantly different from a correlation coefficient of 0 (Holm-Bonferroni corrected $p<0.05$).  Statistical significance was determined using a bootstrap re-sampling with 10,000 re-samplings of each correlation coefficient.  }\label{fig:correlationplots}
\end{figure*}

We now correlate each of the nodal centrality measure described in the previous section with students' homework assignment or exam scores.  
In Fig.~\ref{fig:correlationplots}, we display the results of these calculations.  Statistical significance of the correlation coefficients for each course and assignment type was determined at the $p<0.05$ level using Holm-Bonferroni corrected p-values calculated via a bootstrap re-sampling with 10,000 re-samplings of each correlation coefficient \cite{stats}.  The use of bootstrap re-sampling was motivated by the fact that centrality measures from complex network analysis are inherently interdependent and, thus, violate the assumption of independence fundamental to standard parametric statistics \cite{Brewe2012}.

With respect to our first question of interest --- do well connected students get good grades? --- Fig.~\ref{fig:correlationplots} suggests that this depends both on the type of centrality measure and the course in question.  Four centrality measures have statistically significant correlations to homework grades (both numerical and analytical) for all three courses: closeness centrality, harmonic centrality, in-strength, and out-strength.  Recall that in- and out-strength measure how often a student collaborates by receiving or giving help respectively, while closeness and harmonic centrality are both measures of how ``far'' a student is from other students.   Fig.~\ref{fig:correlationplots} shows that in- and out-disparity have a negative correlation to homework scores in all classes and that correlation is statistically significant for at least one type of homework (analytical or numerical) in two of the three courses.  In- and out-disparity measure a student's tendency to collaborate often with only a small number of students. This result, combined with the significant positive correlations with the four centrality measures above suggest that students who not only collaborate often, but also collaborate significantly with many different people tend to achieve higher grades.  

\begin{figure*} 
\begin{subfigure}[t]{0.5\textwidth}
\includegraphics[page=1, width=\linewidth]{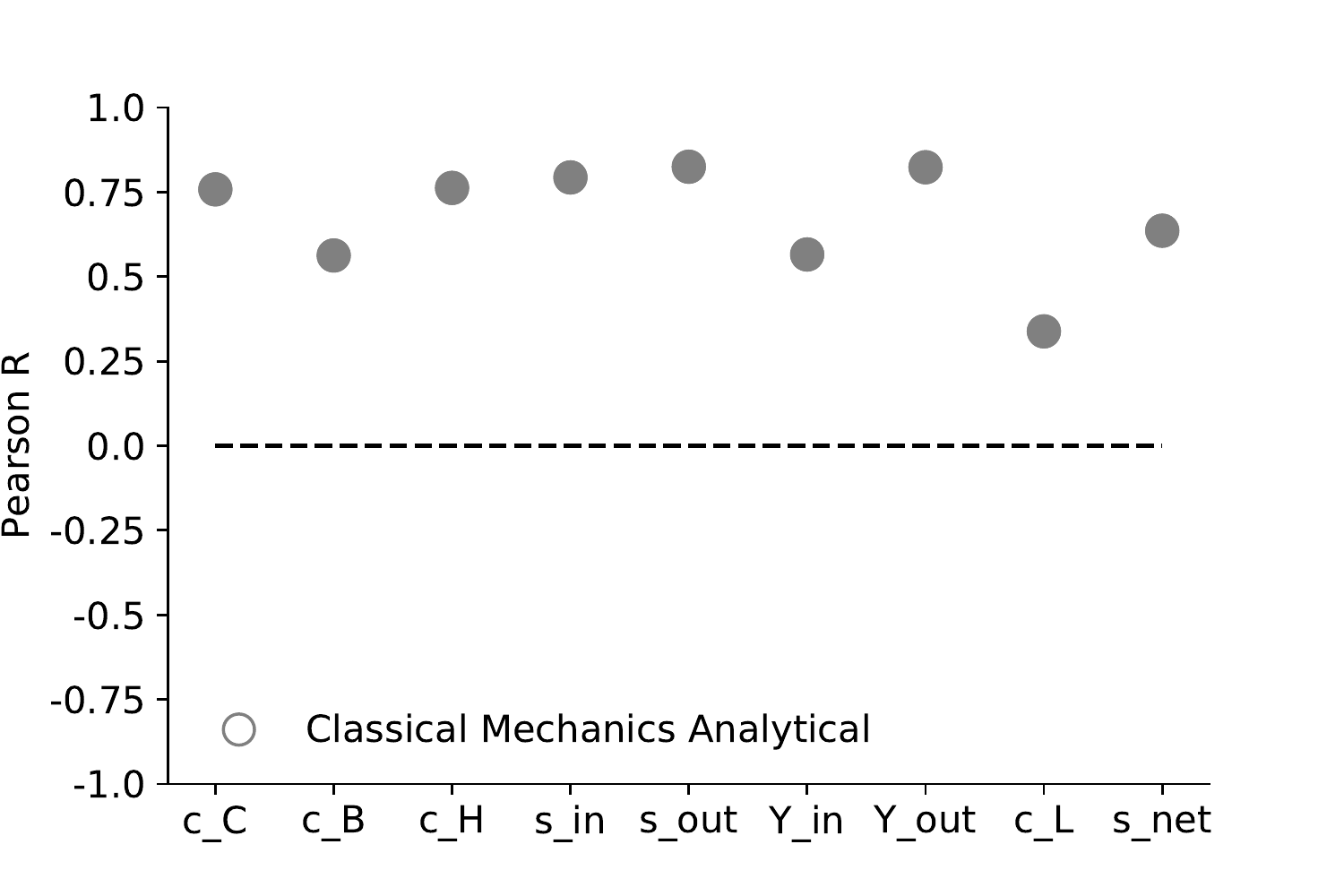}
\label{fig:course_comparisona}
\caption{}
\end{subfigure}%
\begin{subfigure}[t]{0.5\textwidth}
\includegraphics[page=1, width=\linewidth]{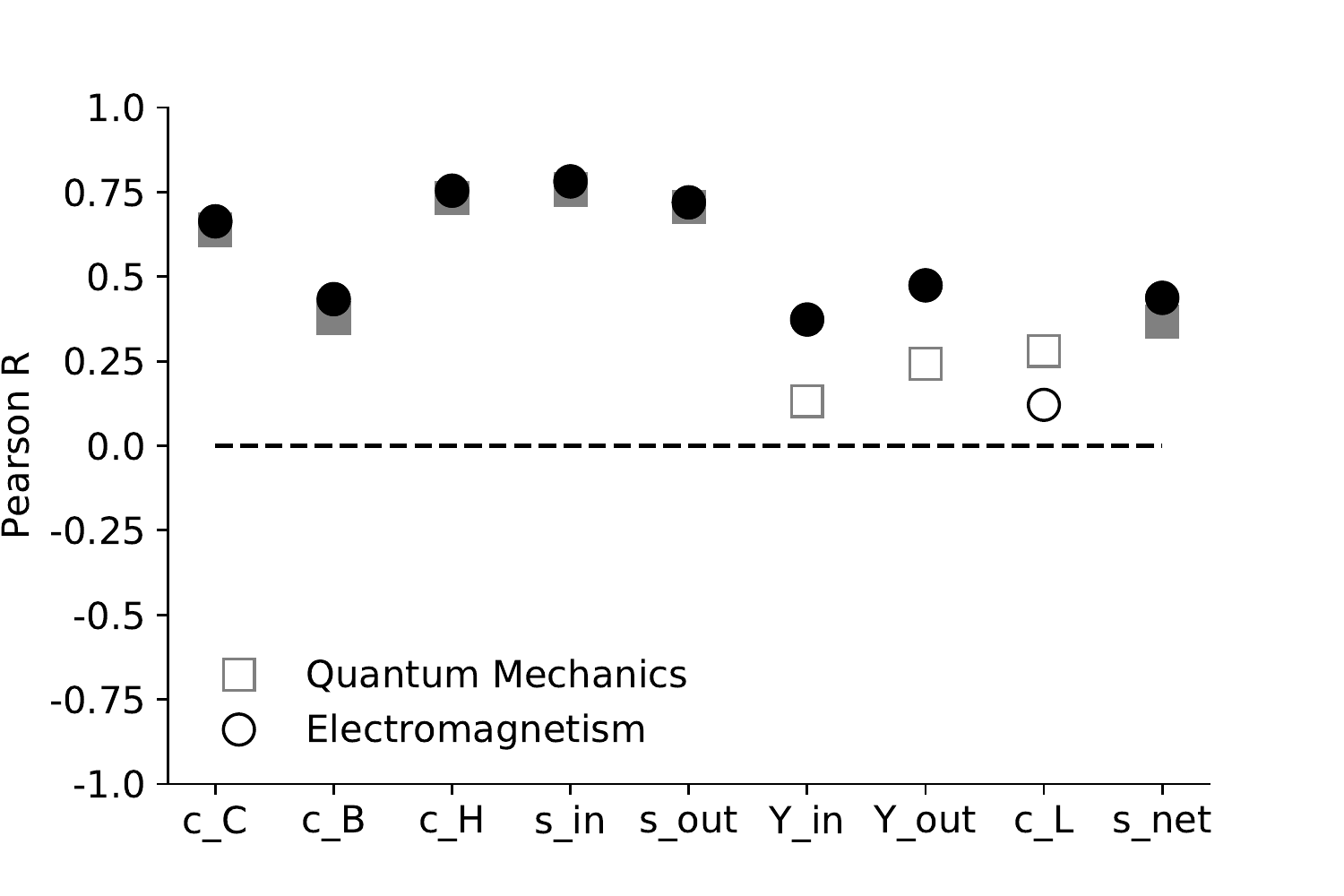}
\label{fig:course_comparisonb}
\caption{}
\end{subfigure}
\caption[Correlation of nodal centrality measures from the Classical Mechanics Analytical network with (a) the corresponding nodal centrality measure from the Classical Mechanics Numerical Network, and (b) the corresponding nodal centrality measure from the other two courses.]{\textit{Correlation of nodal centrality measures from the Classical Mechanics Analytical network with (a) the corresponding nodal centrality measure from the Classical Mechanics Numerical Network, and (b) the corresponding nodal centrality measure from the other two courses: Quantum Mechanics, and Electromagnetism.} Filled markers indicate correlation coefficients that are statistically significantly different from a correlation coefficient of 0 (Holm-Bonferroni corrected $p<0.05$).  Statistical significance was determined using bootstrap re-sampling with 10,000 re-samplings of each correlation coefficient.  }
\label{fig:course_comparison}
\end{figure*}

Fig.~\ref{fig:correlationplots} also shows three measures that tend to have smaller, and less often statistically significant, correlations with homework scores; these are betweenness centrality, local clustering, and net out-strength.  Recall that betweenness centrality measures the extent to which a student can control the flow of information between other students, local clustering determines the extent to which a student may be part of a close knit study group, and net out-strength determines whether a student helps others more than they are helped.  That these measures are less often significant suggests that collaborating often and widely has a greater relation to homework grades than being part of a single close knit study group or being a link between groups or students.  The small correlation of homework scores with net out-strength relative to the correlations for both in-strength and out-strength also suggests that the quantity of collaborations matters more than the type of collaboration (i.e., helping others vs. being helped).  This finding is perhaps surprising given that one might expect that students who more often give help than receive it would be the naturally higher performing students.  

Comparing the correlations in Fig.~\ref{fig:correlationplots} across the three courses shows similar trends across most of the measures.  One significant exception to this is the local clustering coefficient which is larger (and statistically significant) in both Classical Mechanics networks than in the networks for the other two courses.  This suggests that for the Classical Mechanics course, integration into well-established study groups had a stronger relation to students' scores on both the analytical and numerical homework.  To better understand possible sources of this difference, we look at differences between the courses themselves.  Classical Mechanics is the first ``hard'' upper division course most physics students take; in fact, it is generally considered by students to be one of the hardest courses in the Mines physics curriculum.  Thus, students may still be modifying their lower-division study habits to accommodate this additional challenge.  Thus, we may be seeing the effect of tight knit study groups in a way that dissipates as students get to know each other in the ensuing upper-division courses.  The numerical component of the Classical Mechanics course, which was not present in the other courses and is unique to the curriculum, also represents a possible contributing factor.

Another question of interest for this study pertains to whether the benefits of collaboration extend beyond homework scores to students' performance on exams.  It may be that the individual nature of exams suppresses the impact of collaboration on students' performance.  Here again, Fig.~\ref{fig:correlationplots} suggests that the answer depends on what course we look at.  For both Quantum Mechanics and Electromagnetism, correlations between all centrality measures and exam scores are lower than correlations of the same measure with homework scores.  In fact, only out-strength correlates significantly with exam scores and then only in Quantum Mechanics.  This suggests that for these courses, the potential benefit of centrality within the collaboration network does not extend to exams.  On the other hand, in the Classical Mechanics course, the correlations of centrality measures with exam scores is comparable to the corresponding correlations for homework scores for nearly all measures.  Beyond the inclusion of a fairly significant conceptual component in the Classical Mechanics exams ($\sim$1/4 of the questions), it is not obvious why the Classical Mechanics networks appear to correlate better with exam scores.  However, these findings suggest that whether the benefits of collaboration extend to exam scores as well as homework scores depends on the structure of the course and/or content of the exams.  Application of this analysis to other courses will be necessary to pinpoint the course and exam features that best realize the benefits of students' collaboration.  It is worth noting that, depending on the instructional goals of the course, exams may not be designed to realize the benefits of student collaboration.  Moreover, all exams in this study were traditional, individual exams; these findings would likely shift if the format of the exam was less traditional (e.g., group exams).  

In our study, we have two simultaneous networks composed of exactly the same students: the analytical and numerical networks for Classical Mechanics.  This allows us to investigate another question of interest -- how stable are these centrality measures across different types of assignments?  Comparing across the two Classical Mechanics networks (analytical and numerical), we see in Fig.~\ref{fig:correlationplots} nearly identical patterns in the correlations of the different centrality measures with homework scores both in terms of magnitude of the correlations and which correlations are statistically significant.  We also see very similar correlations when correlating network centrality measures created using information from the analytical networks to the scores on the numerical homework and \emph{vice a versa}.  Together these findings suggest that these centrality measures are quite stable across different types of homework.  We can also use these two simultaneous networks to quantify the stability of the roles taken by students in response to different types of homework assignments.  To do this, we correlated centrality measures calculated using the analytical network with those calculated using the numerical network.  We found large correlations (in this case $r>0.5$) for eight of the nine centrality measures (see Fig.\ \ref{fig:course_comparison}a).  Only the local clustering coefficient had only a moderate correlation ($r=0.33$) across networks.  All correlations were statistically significant (Bootstrap re-sampling and Holm-Bonferroni corrected $p<0.05$).  This result suggests that students' collaboration strategies remain relatively stable when presented with different types of homework assignments.

In our study, we also have a large subset of students who took all three courses ($N=67$, see Fig.\ \ref{fig:course_info}).  Focusing specifically on these students, we can also investigate the stability of network centrality measures across time as the students advance from Classical Mechanics to Quantum Mechanics and Electromagnetism.  Since all of these courses are upper-division (typically junior-level) courses, these students have already had much of their undergraduate career to develop collaboration strategies that they believe work for them; thus, we might anticipate that their strategies would be relatively stable over time and across courses.  In Fig.\ \ref{fig:course_comparison}b, we correlate each centrality measure, student-by-student, between Classical Mechanics and Quantum Mechanics and also between Classical Mechanics and Electromagnetism.  Since neither Quantum Mechanics nor Electromagnetism included numerical homework, we utilize the Classical Mechanics analytical network for the purposes of investigating stability between these courses.  Of the nine network centrality measures, four showed large correlations (in this case $r>0.6$, see Fig.\ \ref{fig:course_comparison}b) between the Classical Mechanics network and both the networks of the other two courses.  Interestingly, these are the same four measures with consistent significant correlations with the grades in all three courses (i.e., closeness centrality, harmonic centrality, in-strength, and out-strength).  The remaining centrality measures in Fig.\ \ref{fig:course_comparison}b show smaller correlations suggesting less stability in students' network positions between different course networks.  Local clustering coefficient once again highlights as having a particularly small correlation, which in this case, is not statistically significant.  Thus, the network centrality measures that have the strongest relationship to students' homework scores are also the network centrality measures that appear to be the most stable across networks involving different courses or assignment types.  

\section{Summary and Conclusions}

We utilized the tools of complex network analysis to form social networks based on students' self-reported collaborations when completing regular homework assignments in three upper-division physics courses.  From these networks, we then calculated multiple measures describing the centrality and role of each student within the network.  By correlating these nodal centrality measures with students' scores on both homework and exams, we found that four of the nine centrality measures (closeness centrality, harmonic centrality, in-strength, and out-strength) correlated significantly with students' homework scores in all three classes.  Together, the significance of these four measures suggest that students who not only collaborate often but also collaborate significantly with many different people tend to achieve higher homework grades.  We also found that students' collaboration strategies are relatively stable when presented with different types of homework assignments (e.g., analytical vs. numerical) within the same class.  Finally, we found that while some centrality measures appear to shift significantly when students move into a different course, the four centrality measures most strongly related to students' homework scores are also the most stable between networks from different courses.  Correlations of centrality measures with exam scores were generally smaller than the correlations with homework scores, though this finding varied across courses.  Note that this finding does not necessarily suggest that exams more accurately represent a student's individual understanding; while the correlation with collaboration is generally smaller, the high stakes nature of exams introduces a number of factors besides ability that can impact a student's exam scores (e.g., stereotype threat).  

This work helps provide insight into whether and how students' collaboration impacts their success in the course as measured by course exam and homework grades.  It also contributes to a growing body of research utilizing complex network theory to better understand the role of social networks within the undergraduate classroom.  There are several important limitations to the study.  The findings reported here are correlational and thus cannot clearly establish that broad collaboration improves students' performance, only that students who collaborate broadly tend to have higher scores.  Additionally, these data come from a single institution and have relatively low-N ($N<100$).  Replicating these analyses in additional courses at additional institutions will be important for establishing the generalizability of these findings.  For instructors, these results suggest that encouraging students not only to collaborate, but to collaborate with multiple other students may be an effective strategy towards improving students' homework scores.  Moreover, they suggest that the benefits of collaboration are not automatically transferred to exam performance, but rather the structure of the course and exams can enhance or suppress the relation between student collaboration and exam scores.  Finally, we point out that the same kind of study could be performed on collaborations amongst researchers at the graduate level and beyond, using not only in-class studies in graduate school analogous to those considered here, but also collaboration networks on the arXiv.  A general hypothesis to be examined is whether an overall broader collaboration strategy leads to higher outcomes, in for example h-index, total number of citations, and grant funding.

\section{Acknowledgments}
This material is based in part upon work supported by the US National Science Foundation under grant numbers PHY-1520915, OAC-1740130, and DMR-1407962.  The authors acknowledge the efforts of Esteban Chavez for developing the code that sorted reporting discrepancies in the survey data we collected.  Esteban Chavez was funded under the NSF grant DUE-0836937.  We would like to thank Mark Lusk for sharing course data with us.  Lastly we would like to thank all of the students that participated in our study.  
\bibliography{StudentCollaboration}

\end{document}